\documentclass[a4paper, 11pt]{article}
\usepackage[utf8]{inputenc}
\usepackage[english]{babel}
\usepackage{expl3}
\usepackage{array}
\usepackage{url}
\usepackage{amsmath,amsthm,amsfonts,amssymb,amscd}
\usepackage{fullpage}
\usepackage{algorithm}
\usepackage{algpseudocode}
\usepackage{tabularx}
\usepackage{multirow}
\usepackage{booktabs}
\usepackage{lipsum}
\usepackage{bm}
\usepackage{tikz}
\usetikzlibrary{arrows.meta, positioning}
\usepackage[colorlinks,urlcolor=cobalt,citecolor=cobalt,linkcolor=cobalt,pdftex, pdfauthor = Lukas Exl]{hyperref}
\usepackage{siunitx}
\usepackage{graphicx}
\usepackage{xcolor}
\usepackage{caption}
\usepackage{subcaption}
\usepackage[section]{placeins}
\usepackage{listings}
\usepackage{authblk}
\usepackage{stmaryrd} 

\usepackage{url}

\definecolor{cobalt}{rgb}{0.0, 0.28, 0.67}

\newtheorem{thm}{Theorem}
\newtheorem{lem}[thm]{Lemma}

\newtheorem{rmk}{Remark}

\theoremstyle{definition}

\DeclareUnicodeCharacter{2113}{\ensuremath{\ell}}

\renewcommand{\vec}[1]{\boldsymbol{#1}}

\newcommand*\change[1]{{\color{black}{#1}}}
\newcommand*\changefirst[1]{{\color{black}{#1}}}
\newcommand*\changesecond[1]{{\color{black}{#1}}}

\makeatletter

\newcounter{phase}[algorithm]
\newlength{\phaserulewidth}
\newcommand{\setphaserulewidth}{\setlength{\phaserulewidth}}
\newcommand{\phase}[1]{%
  \vspace{1ex}
  \Statex\strut\refstepcounter{phase}\textbf{Phase~\thephase~--~#1}% Phase text
  }
\makeatother
\setphaserulewidth{.7pt}

\DeclareSymbolFont{largesymbolsA}{U}{txexa}{m}{n}
\DeclareMathSymbol{\varprod}{\mathop}{largesymbolsA}{16}

\date{}

\title{Higher order stray field computation on tensor product domains}
\author[a,b,c]{Lukas Exl \thanks{\texttt{lukas.exl@univie.ac.at}}}
\author[a,b]{Sebastian Schaffer} 

\affil[a]{Wolfgang Pauli Institute, Vienna, Austria}
\affil[b]{Research Platform MMM Mathematics-Magnetism-Materials, University of Vienna, Vienna, Austria}
\affil[c]{Department of Mathematics, University of Vienna, Vienna, Austria} 

\begin{document}
\maketitle
\noindent\textbf{Abstract.}
\noindent We present an extension of the tensor grid method for stray field computation on rectangular domains that incorporates higher-order basis functions. Both the magnetization and the resulting magnetic field are represented using higher-order B-spline bases, which allow for increased accuracy and smoothness. The method employs a super-potential formulation, which circumvents the need to convolve with a singular kernel. The field is represented with high accuracy as a functional Tucker tensor, leveraging separable expansions on the tensor product domain and trained via a multilinear extension of the extreme learning machine methodology. Unlike conventional grid-based methods, the proposed mesh-free approach allows for continuous field evaluation. Numerical experiments confirm the accuracy and efficiency of the proposed method, demonstrating exponential convergence of the energy and linear computational scaling with respect to the multilinear expansion rank.

\noindent\textbf{Keywords.} super-potential, extreme learning machine, operator learning, B-splines, functional Tucker tensor decomposition, mesh-free method, computational micromagnetism\\

\section{Introduction}
Micromagnetic simulations play a crucial role in the study of magnetic materials, particularly in applications such as data storage, sensor systems, and electric motors or generators \cite{suess2018topologically,bashir2012head,fischbacher2018micromagnetics,kovacs2020computational}. Computational approaches in micromagnetism encompass tasks ranging from total energy minimization and modeling the time-dependent evolution of magnetization configurations to solving inverse design problems \cite{exl2020micromagnetism,wang2021inverse,pollok2021inverse}. Developing efficient and reliable numerical methods is therefore of significant interest. Among these tasks, the accurate and efficient computation of the non-local magnetostatic stray field remains the most challenging aspect. Straight-forward implementation of the integral solution operator or the underlying whole space Poisson problem scales quadratically in the number of degrees of freedom, typically related to the amount of nodes or computational cells in a mesh or spatial grid of the magnet domain.   
Current conventional micromagnetic numerical methods are generally classified into non-uniform grid and finite element methods  (FEM) \cite{schrefl2007numerical,chang2011fastmag,abert2013magnum} or uniform grid and finite difference (FD) \cite{miltat2007numerical,donahue1999oommf,vansteenkiste2014design,bjork2021magtense,bruckner2023magnum} approaches, with notable differences in how the computationally demanding stray field is treated. Finite element methods operate on unstructured grids, where local basis functions (typically $P1$) are employed in a Galerkin framework. Stray field computation in this context typically involves solving an interface problem for the scalar potential over the entire space, often handled via a mixed finite element/boundary element (FEM-BEM) approach \cite{hertel2019large,hertel2014hybrid,kakay2010speedup}. The resulting linear systems exhibit sparsity for interior magnetization contributions but are dense for boundary or exterior interactions. To address this, various acceleration techniques have been developed that exploit compression properties based on the separation of near and far field interactions. These include fast multipole methods \cite{visscher2010simple,livshitz2009nonuniform}, hierarchical matrices \cite{hackbusch2009hierarchische,hertel2019large}, and the non-uniform Fast Fourier Transform (NUFFT) \cite{dutt1993fast,potts2001fast}, with applications to micromagnetics introduced in \cite{exl2014non}. Higher order FEM and FEM-BEM methods for magnetic field calculation have been introduced in micromagnetism and magnetohydrodynamics \cite{schrefl2002higher,ilic2003higher,dao2024nodal,sovinec2004nonlinear}.
In contrast, finite difference methods use a structured, equidistant product domain decomposed into uniformly sized cuboidal elements (uniform grids). This regularity enables the use of Fast Fourier Transform (FFT) techniques to accelerate stray field computations \cite{berkov1993solving}. Additionally, certain integrals can be precomputed and stored efficiently due to symmetries and the tensor product structure of the grid \cite{exl2012fast}. Theoretically, all of these methods enable quasi-linear computational scaling in the degrees of freedom and converge up to quadratically in the cell size \cite{abert2013numerical}. Typically, the field is computed via its demagnetization tensor \cite{newell1993generalization,bjork2021magtense} or the scalar potential \cite{abert2011fast} is calculated like in the \textit{tensor grid (TG) method} \cite{exl2012fast} via its convolution integral over the domain $\Omega$, such as
\begin{align}\label{eq:pot1}
    \phi(\vec x) = \frac{1}{4\pi} \int_{\Omega} \vec{m}(\vec{y}) \cdot \frac{\vec{x}- \vec{y}}{|\vec{x} - \vec{y}|^3}\, d\boldsymbol{y},
\end{align}
where the magnetization $\vec{m}$ is assumed to be constant and located at the midpoints of the computational cells, followed by (second order) finite differences for the field computation $\vec h = -\nabla \phi$. Such a midpoint collocation scheme achieves quadratic convergence in the mesh size and a smooth truncation of the singular convolution kernel around the origin can be used, e.g., via Gaussian sums (GS) with cut-off $\varepsilon > 0$ \cite{exl2014fft}. The smoothed convolution method still maintains the convergence rate, and the truncation error of size $\mathcal{O}(\varepsilon)$ only contributes a constant, provided the truncation cut-off $\varepsilon$ is smaller than the cell size. Hence, the field is not affected. The advantage of using a GS approximation of the kernel lies in the coordinate separability owing to the tensor product structure, that is,
\begin{align}
    k_{GS}(|\vec x -\vec y|) = \sum_{s=1}^S \omega_s \prod_{p=1}^3 e^{-\alpha_s (x^{(p)}- y^{(p)})^2}. 
\end{align}
This can be exploited on a Cartesian grid , especially for tensor product structured magnetization \cite{exl2012fast}.
However, positioning the magnetization and the evaluation points at the cell midpoints is critical for attaining quadratic convergence, otherwise, the convergence deteriorates to linear order. 
Hence, the evaluation of the potential and field at arbitrary locations in the domain needs (i) a higher accuracy in the smoothed convolution, i.e., a truncation error as a higher order in $\varepsilon$, together with (ii) a continuous description of the magnetization. 

To address (i) we make use of a \textit{super-potential}:
There holds the following simple relation for the magnetic field 
\begin{align}\label{eq:field_new}
    \vec h (\vec x) = - \nabla \phi(\vec x) \change{\,= \nabla \big(\Delta \big(\nabla \cdot \vec u(\vec x)\big)\big),}
\end{align}
with 
\begin{align}\label{eq:superpot1}
    \vec u(\vec x) = \frac{1}{8 \pi} \int_{\Omega} |\vec x - \vec y| \,\vec m(\vec y)\, d\boldsymbol{y}.
\end{align}
Using a separable GS approximation $k_{GS}$ of the absolute value in the convolution \eqref{eq:superpot1} yields a truncation error of order 4, see Sec.~\ref{sec:method}.
Hence, we can evaluate the components of the super-potential with high precision. After fitting the super-potential onto a B-spline basis, we can take derivatives analytically and project back onto the original basis, see Sec.~\ref{sec:deriv}.

(ii) The continuous description of the magnetization is motivated from low-parametric continuous models of the form $\vec m(\vec x; \vec \omega)$, where the magnetization $\vec m$ is represented as a function of spatial coordinates $\vec x$, parameterized by a relatively small set of variables $\vec \omega$, avoiding the costly discretization used in conventional mesh-based schemes. Already early work in the field based on low-parametric Ritz methods provided foundational insights into magnetic states in small particles and domain walls \cite{kondorsky1979stability, arrott1979micromagnetics, aharoni1971domain}. Nowadays, such models often leverage spectral decompositions of the effective field operator \cite{d2009spectral, perna2022computational} or represent data-driven approaches based on physics-informed neural networks (PINNs) \cite{raissi2019physics,kovacs2022conditional,kovacs2022magnetostatics}, which emerged as promising low-parametric, mesh-free alternatives for micromagnetics and related PDEs in electromagnetism \cite{baldan2023physics,khan2022physics}. Particular emphasis has been placed on data-driven, mesh-free methods for the computationally intensive stray field calculation. Notable contributions include approaches based on physics-informed neural networks and extreme learning machines (ELMs), which enable efficient computation of both the scalar potential and the stray field \cite{khan2019deep,khan2022physics,kovacs2022magnetostatics,beltran2022physics,pollok2023magnetic,schaffer2023physics,schaffer2024constraint}.

In our proposed method we model the magnetization configuration on a (scaled) tensor product domain $\Omega = \varprod_{p=1}^3 [-a_p,a_p] \subseteq [-0.5,0.5]^3$ as a \textit{functional Tucker tensor} \cite{hashemi2017chebfun,dolgov2021functional}, where we use an expansion of multilinear rank-$(r_1,r_2,r_3)$ in a tensor product B-spline basis, see Sec.~\ref{sec:method}. 
The knot vectors defining the 1-dimensional  (locally supported) B-spline basis functions could be interpreted as grid points of a (fictional) tensor product grid - however, the functional description of the magnetization (and the field) allows evaluation at any location in the domain. 
\changesecond{This knot grid is only a formal construct required to define the local support of the B-spline basis and does not constitute a discretization mesh in the conventional sense. Unlike traditional grid-based schemes, where both quantities are tied to the same physical grid, our functional approach decouples the underlying degrees of freedom of the magnetization and field representations.}
Computations on the magnetization, such as differentiation or integration over the tensor product domain, can be performed on the 1-dimensional basis functions, resulting in drastically reduced computational complexity. Micromagnetic algorithms such as total energy minimization can be established within the representation in an analogous way as in ~\cite{exl2014tensor,schaffer2023physics,exl2019optimization}. However, also for a given functional description of a magnetization or a linear combination of the functional Tucker form, it is possible to efficiently compute an accurate approximation of sufficiently large multilinear rank via a \textit{multilinear tensor product extreme learning machine} (ml-tp ELM), see Sec.~\ref{sec:fitting}. 
Finally, also the computation of the magnetostatic energy benefits from the tensor product structure of the magnetization and the field.

In the following, we describe our approach systematically with a brief summary of the algorithm for the stray field computation in the end of the next section. Numerical experiments validate our approach in Sec.~\ref{sec:numerics}

\section{Method}\label{sec:method}
Consider a tensor product domain $\Omega = \varprod_{p=1}^3 [-a_p,a_p] \subseteq [-0.5,0.5]^3$ scaled inside the centered unit cube.
Central to our method is the use of the \textit{super-potential} $u_\rho$ associated with the Newton potential $\mathcal{N}_\rho$ which is induced by a given density $\rho$ \cite{chandrasekhar1989super}. There holds  
\begin{align}\label{eq:newtonpot}
    -\Delta u_\rho(\vec x) = \mathcal{N}_\rho(\vec x) := -\frac{1}{4\pi} \int_{\Omega} \frac{\rho(\vec y)}{|\vec x - \vec y|}\,d\boldsymbol{y}
\end{align}
with 
\begin{align}\label{eq:superpot}
    u_{\rho}(\vec x) = \frac{1}{8 \pi} \int_{\Omega} |\vec x - \vec y| \,\rho(\vec y)\, d\boldsymbol{y}.
\end{align}
The above is a consequence of the fact that $u_\rho$ is a bi-Laplacian potential, while the Newton potential solves Poisson's equation $\Delta \mathcal{N}_\rho = \rho$ \cite{steinbach2007numerical}, that is,
\begin{align}
    \Delta^2 u_\rho = \Delta (\Delta u_\rho) = - \Delta \mathcal{N}_\rho = -\rho. 
\end{align}
Now, for the magnetic scalar potential, we make use of the well-known representation \cite{jackson1999classical}
\begin{align}
    \phi(\vec x) = -\frac{1}{4\pi} \nabla \cdot \int_\Omega \frac{\vec m(\vec y)}{|\vec x - \vec y|} d\boldsymbol{y} = \nabla \cdot \boldsymbol{\mathcal{N}}_{\vec m}(\vec x),
\end{align}
and we arrive at the following simple relation for the magnetic field 
\begin{align}\label{eq:hfield_new_method}
    \boldsymbol{h} (\vec x) = - \nabla \phi(\vec x) = \nabla \big(\nabla \cdot \Delta \vec u_{\vec m}(\vec x)\big) \change{\,= \nabla \big(\Delta \big(\nabla \cdot \vec u_{\vec m}(\vec x)\big)\big)}, 
\end{align}
with $\vec u_{\vec m} = (u_{m^{(1)}},u_{m^{(2)}},u_{m^{(3)}})$, the \textit{vector super-potential} induced by the magnetization components. \change{The last identity in \eqref{eq:hfield_new_method} follows from basic vector calculus and reduces the amount of derivatives since it only involves the scalar Laplacian.}
We use a separable GS approximation $k_{GS}$ of the absolute value in the convolution \eqref{eq:superpot},  that is valid in $(\varepsilon,\delta],\,\delta \geq 1$, which yields a truncation error of order 4. In fact, for $\vec x \in [-\delta/2,\delta/2]^3$ we have (see Lemma~\ref{lem:correctionint})
\begin{align}\label{eq:superpotGS}
     \vec u_{\vec m}(\vec x) = \frac{1}{8 \pi} \int_{\Omega} k_{GS}(|\vec x - \vec y|) \,\vec m(\vec y)\, d\boldsymbol{y} + \mathcal{O}(\varepsilon^4).
\end{align}
Hence, we can evaluate the components of the vector super-potential with high precision, where we use a typical cut-off $\varepsilon$ less than $1$e$-02$.

We model the magnetization configuration on the domain as a \textit{functional Tucker tensor} \cite{hashemi2017chebfun,dolgov2021functional}, where we use an expansion of multilinear rank-$(r_1,r_2,r_3)$ in a tensor product B-spline basis. Each magnetization component $m^{(\mu)}:\,\Omega \rightarrow \mathbb{R},\,\mu=1,2,3$ is represented as 
\begin{equation}\label{eq:mag_bspline}
    {m}^{(\mu)}(\boldsymbol{x}) = \sum_{j_1=1}^{r_1} \sum_{j_2=1}^{r_2} \sum_{j_3=1}^{r_3} \mathcal{A}_{j_1j_2j_3}^{(\mu)} \prod_{p=1}^3 \sigma_{j_p}^{(\mu)}(x^{(p)}),
\end{equation}
where $\mathcal{A}^{(\mu)} \in \mathbb{R}^{r_1 \times r_2 \times r_3}$ defines the \textit{core tensor} of the $\mu$-th magnetization component and the $\sigma_{j_p}^{(\mu)}$ are 1-dimensional B-spline basis functions of order $k>1$ (degree $k-1$) defined on $r_p + k$ knots equidistantly distributed in each direction of the tensor product domain. The functional description of the magnetization (and the field) allows evaluation at any location in the domain. Specifically, the evaluation of ${m}^{(\mu)}(\boldsymbol{x})$ on a tensor grid results in a \textit{Tucker tensor decomposition} \cite{kolda2009tensor}.  Next we show how a given functional description of a magnetization or a linear combination of the form \eqref{eq:mag_bspline} can be accurately and efficiently fitted in the form \eqref{eq:mag_bspline} for given sufficiently large multilinear rank.

\subsection{Fitting a multilinear tensor product ELM}\label{sec:fitting}
The following approach is inspired by the training process of extreme learning machines. In our case, we leverage the multilinear tensor product structure inherent in functional Tucker tensors. This leads to an efficient training algorithm for what we refer to as a \textit{multilinear tensor product extreme learning machine} (ml-tp ELM). 

A conventional ELM is a type of neural network with only a single
hidden layer, where the input data is randomly mapped onto a latent space, and only the output layer requires training \cite{huang2011extreme}. Formally, for an input $\vec{x} \in \Omega$, an ELM with $R$ hidden nodes and a scalar output can be described as
\begin{equation}
   u(\boldsymbol{x};\boldsymbol{a}) = \sum_{j=1}^R a_j \sigma_j(\boldsymbol{x}) \equiv \llbracket \boldsymbol{a}; \boldsymbol{\sigma}(\boldsymbol{x})\rrbracket,
\end{equation}
where $\boldsymbol{a} = (a_j) \in \mathbb{R}^R$ are the trainable parameters and the $\boldsymbol{\sigma} = (\sigma_1,\hdots,\sigma_R):\,\Omega \rightarrow \mathbb{R}^{R}$ comprise the $R$ activation or basis functions. Fitting a function $f$ on $\Omega$ in ELM form can be casted as minimizing the squared $L^2$-loss 
\begin{align}
   \min_{\boldsymbol{a}}\, \int_{\Omega} \big( \llbracket \boldsymbol{a}; \boldsymbol{\sigma}(\boldsymbol{x})\rrbracket - f(\boldsymbol{x})\big)^2\,d\boldsymbol{x}. 
\end{align}
Discretization of the objective function via Monte Carlo or quadrature using $N \geq R$ samples/nodes yields a linear least squares problem
\begin{align}
    \change{\sum_{i=1}^N \omega_i \big( \llbracket \boldsymbol{a}; \boldsymbol{\sigma}(\boldsymbol{x}_i)\rrbracket - f(\boldsymbol{x}_i)\big)^2} = \| H \boldsymbol{a} - \boldsymbol{f} \|^2,
\end{align}
where $H = (H_{ij}) = \big(\sqrt{w_i}\,\sigma_j(\boldsymbol{x}_i)\big) \in \mathbb{R}^{N \times R}$ and $\boldsymbol{f} = (f(\boldsymbol{x}_i)) \in \mathbb{R}^N$. The regularized solution reads ($\alpha > 0$)
\begin{align}
\boldsymbol{a} = \big(H^T H + \alpha I\big)^{-1} H^T \boldsymbol{f}.
\end{align}

Interestingly, there is a \textit{kernel perspective} to ELM. A \textit{kernel ridge regression} (kRR)
\begin{align}
 \min_{\boldsymbol{\xi} \in \mathcal{F}_{\Omega}^{\ast}} 
 \, \sum_{i=1}^N \big( \boldsymbol{\xi} \cdot \psi(\boldsymbol{x}_i) - f(\boldsymbol{x}_i)\big)^2 + \alpha \|\boldsymbol{\xi}\|^2 
\end{align}
with feature map $\psi:\, \Omega \rightarrow \mathcal{F}_{\Omega}$ and kernel function $k(\boldsymbol{x}_1,\boldsymbol{x}_2) = \psi(\boldsymbol{x}_1)\cdot \psi(\boldsymbol{x}_2)$ has an  analogue solution, when considering its low-rank version \cite{schaffer2021machine,exl2021prediction},
\begin{align}
    \boldsymbol{\xi} = \big(\Phi_R^T \Phi_R + \alpha I\big)^{-1} \Phi_R^T \boldsymbol{f}, 
\end{align}
where $\Phi_R \in \mathbb{R}^{N \times R}$ denotes the realization of the feature space map through a low-rank approximation of the kernel matrix. 
This similarity allows for the following conclusion. The hidden layer output of the ELM can be seen as nonlinear embedding in the feature space 
$\mathcal{F}_{\Omega}$. Hence, the basis functions need not to be nonlinear activations, they could be modeled as Gaussians, B-Splines, etc.
defined in $\Omega$.

\change{As basis functions, we choose tensor product B-splines  $\sigma(\boldsymbol{x}) = \sigma_1(x^{(1)})\,\sigma_2(x^{(2)})\,\sigma_3(x^{(3)})$ of order $k>3$ (degree $k-1$) defined on $r_p + k$ knots, i.e., $\bar r^k_p := r_p-k+2$ equidistant knots plus $k-1$ padding knots at each of the boundary points, in each direction of the tensor product domain, resulting in $r_p$ basis functions in each direction.} \changesecond{We choose the basis degree $k-1$ greater or equal than $3$ in order to have smooth first order derivatives for the $L^2$-projection, see Sec.~\ref{sec:deriv}.} 

Our ELM ansatz takes the form of a functional Tucker tensor
\begin{equation}\label{eq:f_bspline}
    u(\boldsymbol{x};\mathcal{A}) = \sum_{j_1=1}^{r_1} \sum_{j_2=1}^{r_2} \sum_{j_3=1}^{r_3} \mathcal{A}_{j_1j_2j_3} \prod_{p=1}^3 \sigma_{j_p}(x^{(p)}) \equiv \big\llbracket \mathcal{A}; \boldsymbol{\sigma}_1 (x^{(1)}), \boldsymbol{\sigma}_2(x^{(2)}),\boldsymbol{\sigma}_3(x^{(3)})\big\rrbracket, 
\end{equation}
where we adopt the tensor bracket short-hand notation from \cite{kolda2009tensor}. Evaluation of $f$ on a tensor product grid, meaning a Cartesian product $\varprod_{p=1}^3 \big\{x_1^{(p)},x_2^{(p)},\hdots,x_{n_p}^{(p)}\big\}$ of grid points along each direction, yields a Tucker tensor decomposition
\begin{align}
     \big\llbracket \mathcal{A}; U_1,U_2,U_3\big\rrbracket := \mathcal{A} \times_1 U_1 \times_2 U_2 \times_3 U_3 \, \in \mathbb{R}^{n_1 \times n_2 \times n_3},
\end{align}
where $\times_p$ denotes the $p$-mode matrix product of the (sparse) \textit{factor matrices} $U_p = \big(\sigma_{j_p}(x_i^{(p)})\big)\in \mathbb{R}^{n_p \times r_p}$ with the \textit{core tensor} $\mathcal{A} \in \mathbb{R}^{r_1 \times r_2 \times r_3}$. We refer to $(r_1,r_2,r_3)$ as the \textit{multilinear rank} of the decomposition \eqref{eq:f_bspline}, and in case of $r \equiv r_p$ simply as the \textit{rank}. 
Fitting a function $f$ defined on the tensor product domain $\Omega$ as a ml-tp ELM leads to the minimization of the residual in the squared $L^2$-norm with respect to the core tensor, that is,
\begin{align}
    \min_{\mathcal{A}} \, \int_{\Omega} \Big( \big\llbracket \mathcal{A}; \boldsymbol{\sigma}_1 (x^{(1)}), \boldsymbol{\sigma}_2(x^{(2)}),\boldsymbol{\sigma}_3(x^{(3)})\big\rrbracket  - f(\boldsymbol{x})\Big)^2\,d\boldsymbol{x}. 
\end{align}
After tensor product quadrature we get a discrete optimization problem of the form 
\begin{align}\label{eq:ml-tp-elm_fitting}
     \min_{\mathcal{A}} \,\Big\|\big\llbracket \mathcal{A}; U_1,U_2,U_3\big\rrbracket - \mathcal{F}\Big\|_F^2. 
\end{align}

\begin{lem}
    Let $U_p \in \mathbb{R}^{n_p \times r_p},\, n_p \geq r_p,\,\, p=1,2,3,$ have full column rank, then the solution to \eqref{eq:ml-tp-elm_fitting} is given as 
    \begin{align}\label{eq:ml-tp-elm_fitting_solution}
        \mathcal{A} = \mathcal{F} \times_1 U_1^\dagger \times_2 U_2^\dagger \times_3 U_3^\dagger.
    \end{align}
\end{lem}
\textit{Proof.} 
Using basic rules for operations on Tucker tensors, we find the derivative of the objective in \eqref{eq:ml-tp-elm_fitting} as
\begin{align*}
    \frac{\partial}{\partial{\mathcal{A}}} \, \big\|\big\llbracket \mathcal{A}; U_1,U_2,U_3 \big\rrbracket - \mathcal{F}\big\|_F^2 =  2 \big( \mathcal{A} \times_1 U_1^TU_1 \times_2 U_2^TU_2 \times_3 U_3^TU_3 - \mathcal{F} \times_1 U_1^T \times_2 U_2^T \times_3 U_3^T \big),
\end{align*}
which vanishes for 
\begin{align}
        \mathcal{A} = \mathcal{F} \times_1 (U_1^TU_1)^{-1} U_1^T \times_2 (U_2^TU_2)^{-1} U_2^T \times_3 (U_3^TU_3)^{-1} U_3^T,
    \end{align}
with $U_p^\dagger = (U_p^TU_p)^{-1} U_p^T,\, p = 1,2,3$, being the pseudo-inverses of the factor matrices.    
\hfill $\square$\\

The pseudo-inverses can be obtained by (regularized) singular value decomposition of the factor matrices of size $n_p \times r_p$. Additionally, the mode multiplications with the pseudo-inverses have to be computed, which has cost $\mathcal{O}(rn^3)$, assuming equal ranks and mode sizes for simplicity. 

\changesecond{Moreover, the tensor-product B-spline basis used in our functional representation is numerically well-conditioned. This property follows from the fact that the inverse (or pseudo-inverse) of the B-spline evaluation matrix enters directly in the fitting step \eqref{eq:ml-tp-elm_fitting_solution}, ensuring stable numerical behavior across different spline orders and ranks.}

If the function $f$ is already represented in the form \eqref{eq:f_bspline} - possibly in a different basis expansion - or as a linear combination of functional Tucker tensors, the corresponding fitting procedure can be directly applied. In this case, the underlying structure can be efficiently exploited to reduce computational complexity. 
For instance, let $\mathcal{F} \in \mathbb{R}^{n_1 \times n_2 \times n_3}$ be given as
\begin{align}\label{eq:bcp}
\mathcal{F} = \big\llbracket \mathcal{B}; V_1,V_2,V_3\big\rrbracket + \big\llbracket \mathcal{C}; W_1,W_2,W_3\big\rrbracket,
\end{align}
with $\mathcal{B} \in \mathbb{R}^{r_1^\prime \times r_2^\prime \times r_3^\prime}$, $\mathcal{C} \in \mathbb{R}^{r_1^{\prime\prime} \times r_2^{\prime\prime} \times r_3^{\prime\prime}}$ and $V_p \in \mathbb{R}^{n_p \times r_p^\prime},\, W_p \in \mathbb{R}^{n_p \times r_p^{\prime\prime}},\,\, p=1,2,3$, 
then the solution for the core tensor $\mathcal{A}$ is
\begin{align}\label{eq:fitting_bcp}
    \mathcal{A} = \mathcal{B} \times_1 \big(U_1^\dagger V_1\big) \times_2 \big(U_2^\dagger V_2\big) \times_3 \big(U_3^\dagger V_3\big)\, + \,\mathcal{C} \times_1 \big(U_1^\dagger W_1\big) \times_2 \big(U_2^\dagger W_2\big) \times_3 \big(U_3^\dagger W_3\big).
\end{align}
The computational cost of forming the factor matrices is $\mathcal{O}\big(n_p r_p (r_p' + r_p'')\big)$, while the cost of performing the mode multiplications is $\mathcal{O}\big(r ({r'}^3 + {r''}^3)\big)$.

\subsection{Derivatives}\label{sec:deriv}
Derivatives of functions in the form \eqref{eq:f_bspline} are carried out by differentiating the corresponding B-spline basis functions. For instance, for a partial derivative with respect to the first coordinate, we have
\begin{align}
    \partial_{x^{(1)}} u(\boldsymbol{x};\mathcal{A}) = \big\llbracket \mathcal{A}; \partial_{x^{(1)}} \boldsymbol{\sigma}_1(x^{(1)}), \boldsymbol{\sigma}_2(x^{(2)}),\boldsymbol{\sigma}_3(x^{(3)})\big\rrbracket.
\end{align}
The degree of the respective B-spline basis functions is decreased by one.
However, we fit the derivative $ \partial_{x^{(p)}} u(\boldsymbol{x};\mathcal{A}) $ onto the original basis using the \changesecond{$L^2$-projection} procedure described in the previous section. 
\changesecond{For smooth data, the $L^2$-projection is stable and retains the expected convergence order for derivatives; the only effect is a slightly smoother representation at the knots than the native derivative, which is immaterial for our purposes.}
This approach allows higher-order derivatives to be computed through successive applications of first-order differentiation each followed by $L^2$-projection \changesecond{avoiding assembling higher-order derivative operators.} \changesecond{Our order assumption $k>3$ (i.e., degree at least $3$) ensures that the first derivative is continuously differentiable.} \changesecond{This "differentiate $\rightarrow$ project" strategy controls the growth of operator norms that typically accompanies higher-order differentiation and preserves the optimal asymptotic rates for smooth targets, while keeping all quantities in a single space.} Differential operators involving sums of partial derivatives - such as the divergence or Laplacian - are handled using the same fitting strategy, applied to linear combinations of functional Tucker tensors.  

\subsection{Separable approximation of the super-potential}\label{sec:GS}
In order to exploit the coordinate separability of the tensor product basis functions, we establish an approximation of the $|\boldsymbol{x}|$-function which is valid in an interval $(\varepsilon,\delta],\,\delta \geq 1$. 

\begin{figure}
\center
\includegraphics[width=0.8\textwidth]{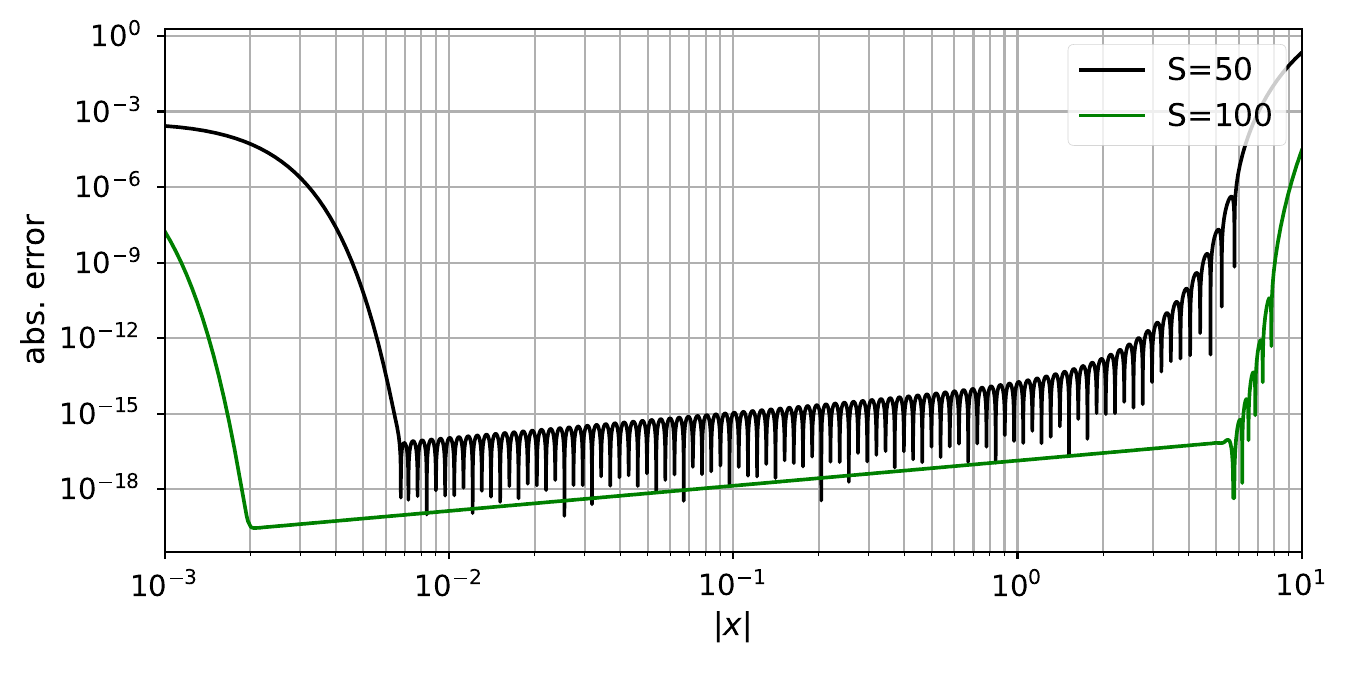}
\caption{\change{Absolute errors of the GS approximation for $|x|$ with $S=\change{50}$ and $S=100$ terms}.}
\label{fig:GSapprox}
\end{figure}

We start from the representation formula 
\begin{align}
 \int_0^\infty x^n e^{-\rho x^2}\,\text{d}x = \frac{\Gamma (\tfrac{n+1}{2})}{2\rho^{\tfrac{n+1}{2}}},\,\, \rho >0,\, n>-1.
\end{align}
We choose $n=0$, and following \textit{sinc quadrature} \cite{hackbusch2006low} this yields for $\rho=|\boldsymbol{x}|^2$ to an exponentially convergent separable representation of $1/|\boldsymbol{x}|$, that is, for $0<\varepsilon<1\leq\delta$
\begin{align}\label{eq:sin_1_x}
    \frac{1}{|\boldsymbol{x}|} \approx \sum_{s=1}^S \omega_s \, e^{-\alpha_s |\boldsymbol{x}|^2} = \sum_{s=1}^S \omega_s\, \prod_{p=1}^3 e^{-\alpha_s (x^{(p)})^2}, \, |\boldsymbol{x}| \in (\varepsilon,\delta].
\end{align}
Multiplying \eqref{eq:sin_1_x} by $|\boldsymbol{x}|^2$ gives
\begin{align}\label{eq:gs_x}
    |\boldsymbol{x}| \approx k_{GS}(|\boldsymbol{x}|) = \sum_{s=1}^S \omega_s \,|\boldsymbol{x}|^2\, e^{-\alpha_s |\boldsymbol{x}|^2} = \sum_{s=1}^S \sum_{q=1}^3 \omega_s 
\left( \prod_{p=1}^3 
\begin{cases}
(x^{(p)})^2\, e^{-\alpha_s (x^{(p)})^2}, & \text{if } p = q \\[0.5ex]
e^{-\alpha_s (x^{(p)})^2}, & \text{if } p \neq q
\end{cases}
\right),
\end{align}
which is a sum of rank-3 separable functions. The sinc quadrature can be adapted in terms of error and valid interval \cite{exl2014tensor}. Fig.~\ref{fig:GSapprox} shows the absolute errors of a GS approximation \change{with $50$ and $100$ terms.} Using $k_{GS}(|\boldsymbol{x}|)$ instead of the kernel $|\boldsymbol{x}|$ in the super-potential calculation \eqref{eq:superpot} leads to an error of $\mathcal{O}(\varepsilon^4)$, under the assumption that the difference $\big||\vec x| - k_{GS}(|\vec x|)\big|$ is negligible for $|\vec x| \in (\varepsilon,\delta]$. That is, we assume for the correction integral
\begin{align}
    \int_\Omega  \big(|\vec x-\vec y|- k_{GS}(|\vec x - \vec y|)\big) \,\rho(\vec y)\, d\boldsymbol{y} \approx   \int_{\mathcal{B}_{\varepsilon}(\vec x)}  \big(|\vec x-\vec y|- k_{GS}(|\vec x - \vec y|)\big) \,\rho(\vec y)\, d\boldsymbol{y},
\end{align}
with $\mathcal{B}_{\varepsilon}(\vec x)$ denoting the ball of radius $\varepsilon$ centered at $\boldsymbol{x}$.
\begin{lem}\label{lem:correctionint}
Let $\rho$ be bounded. Then there holds 
\begin{align}
     \Big|\int_{\mathcal{B}_{\varepsilon}(\vec x)} \big(|\vec x-\vec y|- k_{GS}(|\vec x - \vec y|)\big) \,\rho(\vec y)\, d\boldsymbol{y} \Big| = \mathcal{O}(\varepsilon^4).
\end{align}
\end{lem}
\textit{Proof.} For $\boldsymbol{y} \in \mathcal{B}_{\varepsilon}(\vec x)$ we have $|\vec x-\vec y| \leq \varepsilon$ and $k_{GS}(|\vec x - \vec y|) \leq C \varepsilon^2 + \mathcal{O}(\varepsilon^4)$. Hence, we have 
\begin{align}
    \Big|\int_{\mathcal{B}_{\varepsilon}(\vec x)} \big(|\vec x-\vec y|- k_{GS}(|\vec x - \vec y|)\big) \,\rho(\vec y)\, d\boldsymbol{y} \Big| \leq \varepsilon \,\big(1+C\, \varepsilon\, + \mathcal{O}(\varepsilon^3)\big) \, |\mathcal{B}_{\varepsilon}(\vec x)| \, \|\rho\|_\infty= \mathcal{O}(\varepsilon^4), 
\end{align}
with $|\mathcal{B}_{\varepsilon}(\vec x)| = \mathcal{O}(\varepsilon^3)$ the volume of the $\varepsilon$-ball.
\hfill $\square$

\begin{rmk}\label{rmk:superpoterror} 
\change{Setting $\rho(\vec y) = 1$ and integrating $|\int_{\mathcal{B}_{\varepsilon}(\vec x)} (|\vec x-\vec y|- k_{GS}(|\vec x - \vec y|))\, d\boldsymbol{y}|$ analytically over $\mathcal{B}_{\varepsilon}$ yields an upper bound of the absolute error. From Fig.~\ref{fig:GSapprox} we have $\varepsilon \approx \num {2e-03}$ which yields an error $\leq \num{3.3e-14}$ for the super-potential.} 
\end{rmk}

When using \eqref{eq:gs_x} the approximation of a component of the super-potential \eqref{eq:superpotGS} induced by a magnetization component $m^{(\mu)}$ of the form \eqref{eq:f_bspline} gets a sum of functional Tucker tensors, that is,
\begin{align}\label{eq:superpot_approx}
    u_{m^{(\mu)}}(\boldsymbol{x}) \approx \frac{1}{8\pi}\sum_{s=1}^S \omega_s \sum_{q=1}^3 \sum_{j_1,j_2,j_3} \mathcal{A}_{j_1j_2j_3}^{(\mu)} \, \iota_{j_q,s}^{(q)}(x^{(q)})  \prod_{p \neq q} \iota_{j_p,s}^{(p)}(x^{(p)}), 
\end{align}
with the integrals 
\begin{eqnarray}\label{eq:ints}
\begin{aligned}
\iota_{j_q,s}^{(q)}(x^{(q)}) &= \int_{-a_q}^{a_q} (x^{(q)} - y^{(q)})^2\, e^{-\alpha_s (x^{(q)} - y^{(q)})^2} \sigma_{j_q}(y^{(q)})\, dy, \\
\iota_{j_p,s}^{(p)}(x^{(p)}) &= \int_{-a_p}^{a_p} e^{-\alpha_s (x^{(p)} - y^{(p)})^2} \sigma_{j_p}(y^{(p)})\, dy, \,\,\,(p \neq q).
\end{aligned}
\end{eqnarray}

The integrals can be precomputed up to machine precision using adaptive Gauss-Kronrod quadrature for given target locations $\boldsymbol{x}$ on a tensor grid. For micromagnetic energy calculations, the super-potential is required within the domain. However, it can also be evaluated on an extended tensor grid, where the GS approximation remains accurate.

\subsection{Field and energy calculation}\label{sec:field}
We apply the fitting procedure \eqref{eq:fitting_bcp} on the separable approximation of the super-potential components to represent it as functional Tucker tensor with multilinear rank $(r_1^\prime,r_2^\prime,r_3^\prime)$ of the form
\begin{align}
    u_{m^{(\mu)}}(\boldsymbol{x}) \approx \big\llbracket \mathcal{A}_{u^{(\mu)}} ; \boldsymbol{\sigma}_1 (x^{(1)}), \boldsymbol{\sigma}_2(x^{(2)}),\boldsymbol{\sigma}_3(x^{(3)})\big\rrbracket.
\end{align}
Evaluation of \eqref{eq:superpot_approx} on a tensor grid of  $n_p\, (\geq r_p^\prime)$ Gaussian quadrature target points in each spatial direction results in a sum of Tucker tensors (cf. \eqref{eq:bcp})
\begin{align}\label{eq:bcp_field}
     \frac{1}{8\pi} \sum_{s=1}^S \omega_s \big(\big\llbracket \mathcal{A}^{(\mu)}; I^{(1)}_{1,s},I^{(1)}_{2,s},I^{(1)}_{3,s}\big\rrbracket + \big\llbracket \mathcal{A}^{(\mu)}; I^{(2)}_{1,s},I^{(2)}_{2,s},I^{(2)}_{3,s}\big\rrbracket  + \big\llbracket \mathcal{A}^{(\mu)}; I^{(3)}_{1,s},I^{(3)}_{2,s},I^{(3)}_{3,s}\big\rrbracket \big).
\end{align}
Note, that we have $I^{(2)}_{1,s} = I^{(3)}_{1,s},\, I^{(1)}_{2,s} = I^{(3)}_{2,s}$ and $I^{(1)}_{3,s} = I^{(2)}_{3,s}$.

The least squares solution for $\mathcal{A}_{u^{(\mu)}} \in \mathbb{R}^{r_1^\prime \times r_2^\prime \times r_3^\prime}$ is therefore
\begin{align}\label{eq:superpotcore}
   \mathcal{A}_{u^{(\mu)}} =  \frac{1}{8\pi} \sum_{s=1}^S \omega_s \sum_{q=1}^3  \mathcal{A}^{(\mu)} \times_1 \big(U_1^\dagger\, I^{(q)}_{1,s}\big) \times_2 \big(U_2^\dagger\, I^{(q)}_{2,s}\big) \times_3 \big(U_3^\dagger\, I^{(q)}_{3,s}\big),
\end{align}
with $U_p^\dagger \in \mathbb{R}^{r_p^\prime \times n_p},\,p=1,2,3$.

The field $\boldsymbol{h} (\vec x) = \change{\,\nabla \big(\Delta \big(\nabla \cdot \vec u_{\vec m}(\vec x)\big)\big)}$ 
is now obtained 
by applying successive first-order differentiation  on the super-potential components in functional Tucker form.

The stray field energy is 
\begin{align}
e = -\frac{1}{2} \int_\Omega \boldsymbol{h}(\boldsymbol{x}) \cdot \boldsymbol{m}(\boldsymbol{x})\,d\boldsymbol{x},
\end{align}
which gets a sum of three inner products of Tucker tensors \cite{bader2008efficient} when approximated with a tensor product quadrature, e.g., Gaussian quadrature.  

\change{\subsection{Operator learner perspective and algorithm summary}}
\change{The core step, fitting $\mathbf{u}_m$ from $\mathbf{m}$ modeled as B-spline functional Tucker tensors, can be seen as learning a low-rank surrogate of the linear stray-field operator. Concretely, we use a $L^2$-projector onto \emph{multilinear tensor-product Extreme Learning Machines} (ml-tp ELM): the B-spline factors act as fixed feature maps along each axis, and only the smaller Tucker core(s) are solved by (regularized) least squares via pseudo-inverses, which links directly to low-rank kernel ridge regression. This treats the convolution-to-super-potential map as an \emph{operator learner}: the separable Gaussian-sum (GS) kernel supplies a data-efficient, physics-informed featurization, and the learned core provides an adaptive compression of the operator’s action. After this fit, the remaining differential operators are applied analytically in the same space, yielding a scalable, higher-order, mesh-free pipeline from $\mathbf{m}$ to $\mathbf{h}$ and the energy.
Working with the vector super-potential $\mathbf{u}_{\vec m}$, whose components are
\[
u_{m^{(\mu)}}(\vec x)=\frac{1}{8\pi}\int_{\Omega}\!\lvert \vec x- \vec y\rvert\, m^{(\mu)}(\vec y)\,d\vec y,\qquad \mu=1,2,3,
\]
turns the singular $1/\lvert \vec x- \vec y\rvert$ convolution for the scalar potential into a \emph{smoother} kernel, and the field is recovered via the identity
\[
\mathbf{h}(\vec x)=\nabla\!\big(\Delta(\nabla\!\cdot \mathbf{u}_{\vec m}(\vec x))\big).
\]
This (i) removes singular integrals, (ii) reduces derivative complexity to first derivatives plus the scalar Laplacian, enabling the stable ``differentiate $\to$ project'' procedure in the B-spline/Tucker space, and (iii) admits a separable GS approximation of $\lvert \vec x-\vec y\rvert$ with truncation error $O(\varepsilon^4)$, which is crucial for accurate, fast, low-rank tensor computations on tensor-product domains.}

A concise summary of the field and energy computation procedure is provided in Alg.\ref{alg:ho-stray-field}.

\begin{algorithm}
\caption{Higher order stray field and energy computation on a tensor product domain}
\label{alg:ho-stray-field}
\begin{algorithmic}
\Require Components of the magnetization $\vec m$ in functional Tucker form \eqref{eq:mag_bspline} with multilinear rank $(r_1,r_2,r_3)$. Multilinear rank $(r_1^\prime,r_2^\prime,r_3^\prime)$ for super-potential (and field) components. Mode sizes $n_p > r_p$ (e.g. $n_p = 2r_p$) for ml-tp ELM fitting. Order $k>1$ of B-spline basis.
\phase{Precomputation:}
\State Setup GS approximation.
\State Precompute the matrices $I_{p,s}^{(q)} \in \mathbb{R}^{n_p \times r_p},\, p,q = 1,2,3; s=1,\hdots,S$ (s. \eqref{eq:ints}, \eqref{eq:bcp_field}).
\State Precompute pseudo-inverses of $U_p \in \mathbb{R}^{n_p \times r_p^\prime}$ and matrix products with $I_{p,s}^{(q)}$ in \eqref{eq:superpotcore}. 
\phase{Computation of the field $\boldsymbol{h}(\boldsymbol{x})$:}
\State Compute core tensors $\mathcal{A}_{u^{(\mu)}},\, \mu = 1,2,3$ (s. \eqref{eq:superpotcore}). 
\State Compute components of field $\boldsymbol{h} (\vec x) = \change{\,\nabla \big(\Delta \big(\nabla \cdot \vec u_{\vec m}(\vec x)\big)\big)}$ according to Sec.~\ref{sec:deriv} as functional Tucker tensors.
\phase{Computation of energy}
\State Compute energy $e$ as sum of inner products of the functional Tucker tensors of the field and magnetization components via tensor product Gaussian quadrature.
\end{algorithmic}
\end{algorithm}

\subsection{Storage and computational costs}
In the precomputation we need to store $6\times S$ matrices with a total storage requirement of $2 S \sum_{p=1}^3 n_p r_p$, plus two vectors of length $S$ for the GS approximation. 
The magnetization components as well as the field in functional Tucker needs the storage of their core tensors, which amounts to $\prod_{p=1}^3 r_p$ and $\prod_{p=1}^3 r_p^\prime$, respectively. 
In the course of Alg.~\ref{alg:ho-stray-field} we need the knot vectors  of length $r_p+k$ and $r_p^\prime+k,\,p=1,2,3$, to construct the basis functions, as well as the matrices $U_p$ of size $n_p \times r_p^\prime, \, p=1,2,3$ or their pseudo inverses. 

The computation of the core tensor \eqref{eq:superpotcore} consists of three SVDs for the pseudo inverses of the factor matrices $U_p$, their matrix products with the matrices $I_{p,s}^{(q)}$ with cost $\mathcal{O}\big(n_p r_p r_p^\prime\big)$ each and $3S$ tensor times matrix in all modes with the core tensors of the magnetization components. The first two steps can be precomputed, while the latter costs $\mathcal{O}\big(S \sum_{p=1}^3 r_p^3r_p^\prime\big)$ to compute, assuming equal multilinear rank for all magnetization components \cite{bader2008efficient}. 
Derivatives, such as Laplacian, divergence and gradient, have a computational cost comparable to that of fitting a ml-tp ELM of a (sum of) functional Tucker tensors, which is similar to the computation of the super-potential core tensor.  
The energy computation costs $\mathcal{O}\big(\sum_{p=1}^3 n_pr_p r_p^\prime + r_p^3 r_p^\prime\big)$, if we assume $r_p^\prime \leq r_p$. If $r_p^\prime  > r_p$, the roles of $r_p$ and $r_p^\prime$ are interchanged accordingly  \cite{bader2008efficient}.

The dense version of the tensor grid method (TG) \cite{exl2012fast} for stray field calculation (i.e., without compression) can be viewed as a special case of the higher-order scheme presented here, corresponding to the case $k=1$, though it is formulated as a finite difference method and does not employ the super-potential approach. The computational scaling of both approaches becomes comparable when the tensor ranks are equal to the grid size used in the TG method. Similarly, other common finite difference methods can be compared in terms of storage requirements and computational scaling under these conditions \cite{abert2013numerical}. However, the use of a higher order in our approach enables accurate results with relatively small ranks. 

%\newpage
\section{Numerical validation}\label{sec:numerics}
\change{For the numerical experiments we used our JAX implementation \cite{code} and an NVIDIA Quadro GV100 in double machine precision. The \texttt{mumax$^3$} \cite{vansteenkiste2014design} computations used the same hardware and the TG method \cite{exl2012fast} computations were done on a notebook with Intel Core i7-8565U.}
\changesecond{In order to compare traditional mesh-based methods with our functional approach, we refer the degrees of freedom of these methods with the number of equidistantly chosen knot points $\bar r_k^{(p)}$ defined in Sec.~\ref{sec:fitting} in the B-spline expansion.}

\subsection{Fitting magnetization configurations}
To illustrate the fitting procedure described in Sec.~\ref{sec:fitting}, we apply it to various magnetization states previously used to benchmark different stray field computation methods, including the tensor grid (TG) method \cite{abert2013numerical}.

We estimate the generalization error in the maximum norm including all three magnetization components, evaluated on a 
\change{test grid $\mathcal{S}_{\text{test}}$ of $200\times200\times200$ equidistant points.
The error is defined as
\begin{equation}
    \mathrm{Err} = \max_{\mu=1,2,3} \max_{\boldsymbol{x} \in \mathcal{S}_{\text{test}}} \left| m^{(\mu)}(\boldsymbol{x}) - m^{(\mu)}_{\text{appr}}(\boldsymbol{x}) \right|,
\end{equation}
}
where the magnetization components \( m^{(\mu)} \) are approximated by a functional Tucker tensor \( m^{(\mu)}_{\text{appr}} \) of rank \( r \equiv r_p \), constructed using a basis of order \( k \). Throughout the test examples, we used $n_p \equiv 140$.

%-------------  Tables for flower and vortex fitting example ----------------

\begin{table}[h!]
\caption{\changefirst{Max. errors including all magnetization components and timings for fitting a flower state ($a=c=1, b=2$) and a vortex state ($r_c=0.14$) in a unit cube. B-spline order \(k\), rank \(\bar r^k \equiv \bar r^k_p\), max. errors \(Err\), and timings \(t\) in sec.}}
\label{tab:flower_vortex_fitting}
\centering
\setlength{\tabcolsep}{16pt}
\changefirst{
\begin{tabular}{c c c c c}
\toprule
\(k\) & \(\bar r^k\) & $\mathrm{Err}$ flower & $\mathrm{Err}$ vortex & $t$ $[s]$ \\
\midrule
3 &  10  & \num{1.259e-05} & \num{1.068e-01} & \num{3.5e-03} \\
3 &  20  & \num{1.320e-06} & \num{1.143e-02} & \num{4.5e-03} \\
3 &  40  & \num{1.521e-07} & \num{9.800e-04} & \num{7.8e-03} \\
3 &  80  & \num{2.299e-08} & \num{1.560e-04} & \num{1.3e-02} \\
\midrule
4 &  10  & \num{4.410e-07} & \num{1.905e-01} & \num{3.2e-03} \\
4 &  20  & \num{2.167e-08} & \num{5.038e-03} & \num{4.0e-03} \\
4 &  40  & \num{1.284e-09} & \num{1.609e-04} & \num{7.8e-03} \\
4 &  80  & \num{1.248e-10} & \num{1.210e-05} & \num{1.4e-02} \\
\midrule
5 &  10  & \num{2.158e-08} & \num{7.292e-02} & \num{3.3e-03} \\
5 &  20  & \num{4.909e-10} & \num{1.583e-03} & \num{4.1e-03} \\
5 &  40  & \num{1.366e-11} & \num{1.763e-05} & \num{8.9e-03} \\
5 &  80  & \num{6.677e-13} & \num{6.489e-07} & \num{1.4e-02} \\
\midrule
6 &  10  & \num{1.718e-09} & \num{1.300e-01} & \num{3.4e-03} \\
6 &  20  & \num{1.715e-11} & \num{1.128e-03} & \num{4.3e-03} \\
6 &  40  & \num{2.361e-13} & \num{3.227e-06} & \num{7.8e-03} \\
6 &  80  & \num{1.466e-14} & \num{5.203e-08} & \num{1.4e-02} \\
\midrule
7 &  10  & \num{1.407e-10} & \num{5.147e-02} & \num{4.3e-03} \\
7 &  20  & \num{5.844e-13} & \num{4.472e-04} & \num{5.2e-03} \\
7 &  40  & \num{1.987e-14} & \num{5.676e-07} & \num{8.4e-03} \\
7 &  80  & \num{1.588e-14} & \num{4.041e-09} & \num{1.4e-02} \\
\bottomrule
\end{tabular}
}
\end{table}

\changefirst{Tab.~\ref{tab:flower_vortex_fitting} shows errors and timings for varying order and rank in the case of the flower and the vortex state in the unit cube. For the rank we use the number of equidistant knots excluding the padding knots defined as $\bar r^k_p := r_p-k+2$, whereas it is chosen the same in each direction, i.e., $\bar r^k \equiv \bar r^k_p$. 
}

As expected, the error decreases with increasing B-spline order and rank, while the computational cost grows approximately linearly with the rank.
\changesecond{These results clearly illustrate the tradeoff between spline order and the number of basis functions. In particular, they emphasize how increasing the spline order can compensate for a coarser knot grid, effectively achieving a form of numerical compression.}

\subsection{Fitting the super-potential}
Next we estimate the error of the separable approximation \eqref{eq:superpot_approx} of the super-potential, with the GS approximation illustrated in Fig.~\ref{fig:GSapprox} \change{with $S=100$}, in the case of the flower state magnetization fitted as functional Tucker tensor with varying order and rank \change{$\bar r^k_p \equiv 40$}. We use the same basis order for the magnetization and the super-potential components. 
\change{We use a $10\times10\times10$ equidistant grid for the unit cube (including the boundary)}
and calculate the maximum absolute error for all three magnetization components relative to a direct adaptive Gauss-Kronrod quadrature for \eqref{eq:superpot} with absolute tolerance of $1$e$-14$ and the ground truth magnetization, see Tab.~\ref{tab:superpot_errors}. The errors confirm our reasoning from Remark~\ref{rmk:superpoterror}. 

%-------------  Tables for flower example2 ----------------
\begin{table}[h!]
\caption{\change{Max. errors for calculating the separable approximation \eqref{eq:superpot_approx} for the flower state in a unit cube. B-spline order \(k\), rank $\bar{r}^k \equiv \bar r^k_p$ and max. errors \(\mathrm{Err}\). The error was computed on a $10\times10\times10$ equidistant grid.}}
\label{tab:superpot_errors}
\centering
\setlength{\tabcolsep}{16pt}
\change{
\begin{tabular}{c c c}
\toprule
\(k\) & \(\bar r^k\) & \(\mathrm{Err}\) \\
\midrule 
2 &  40  & \num{9.458e-08}\\
3 &  40  & \num{1.012e-10}\\
4 &  40  & \num{7.760e-14}\\
5 &  40  & \num{2.931e-14}\\
6 &  40  & \num{2.697e-14}\\
\bottomrule
\end{tabular}
}
\end{table}

\subsection{Field and energy computation}
\change{In this section we compute the field and the enrgy for various setups, where we use $S=100$ and order $k>3$. The results are compared with those of the DM method \cite{abert2013numerical},  \texttt{mumax$^3$} \cite{vansteenkiste2014design} and the original TG method \cite{exl2012fast}}.

\subsubsection{\change{Homogeneously magnetized cube}}
\begin{figure}
\center
\includegraphics[width=0.75\textwidth]{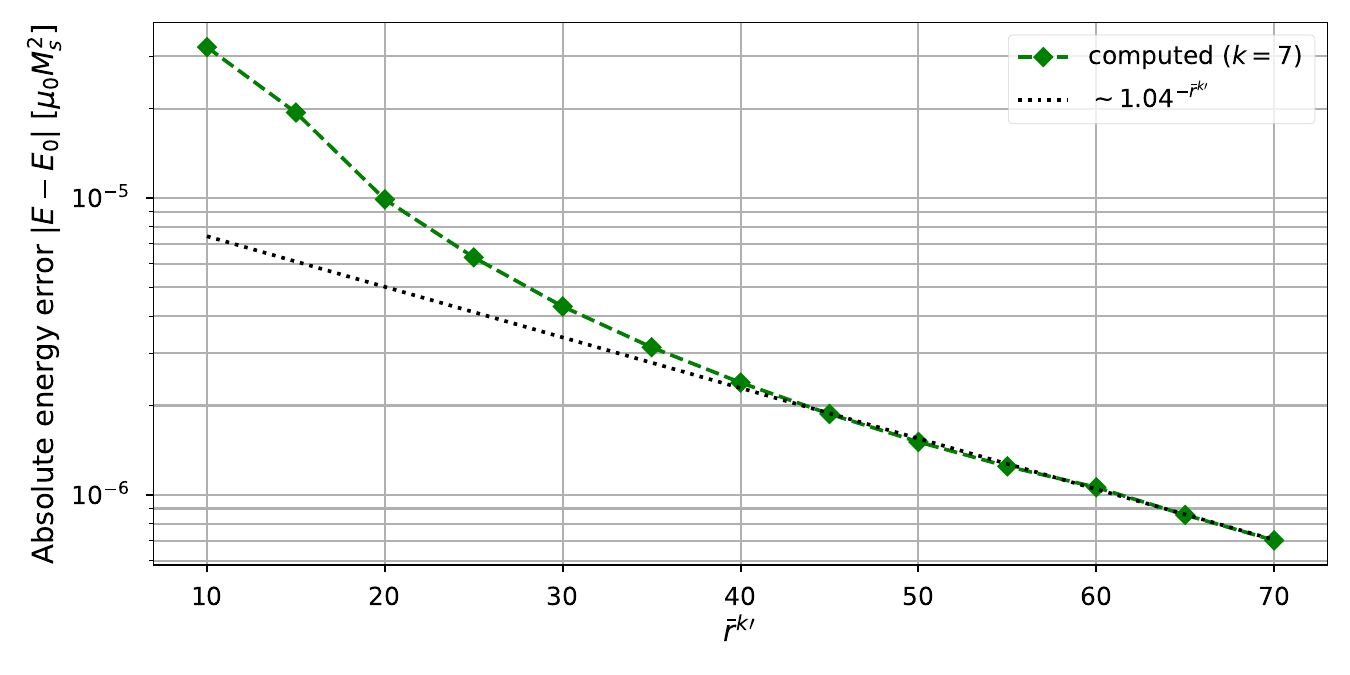}
\caption{\label{fig:energy_uniform} \change{Errors of the energy calculated with Alg.~\ref{alg:ho-stray-field} for a homogeneously magnetized cube with $S=100$ and order $k=7$.}}
\end{figure}

\begin{figure}
\center
\includegraphics[width=0.95\textwidth]{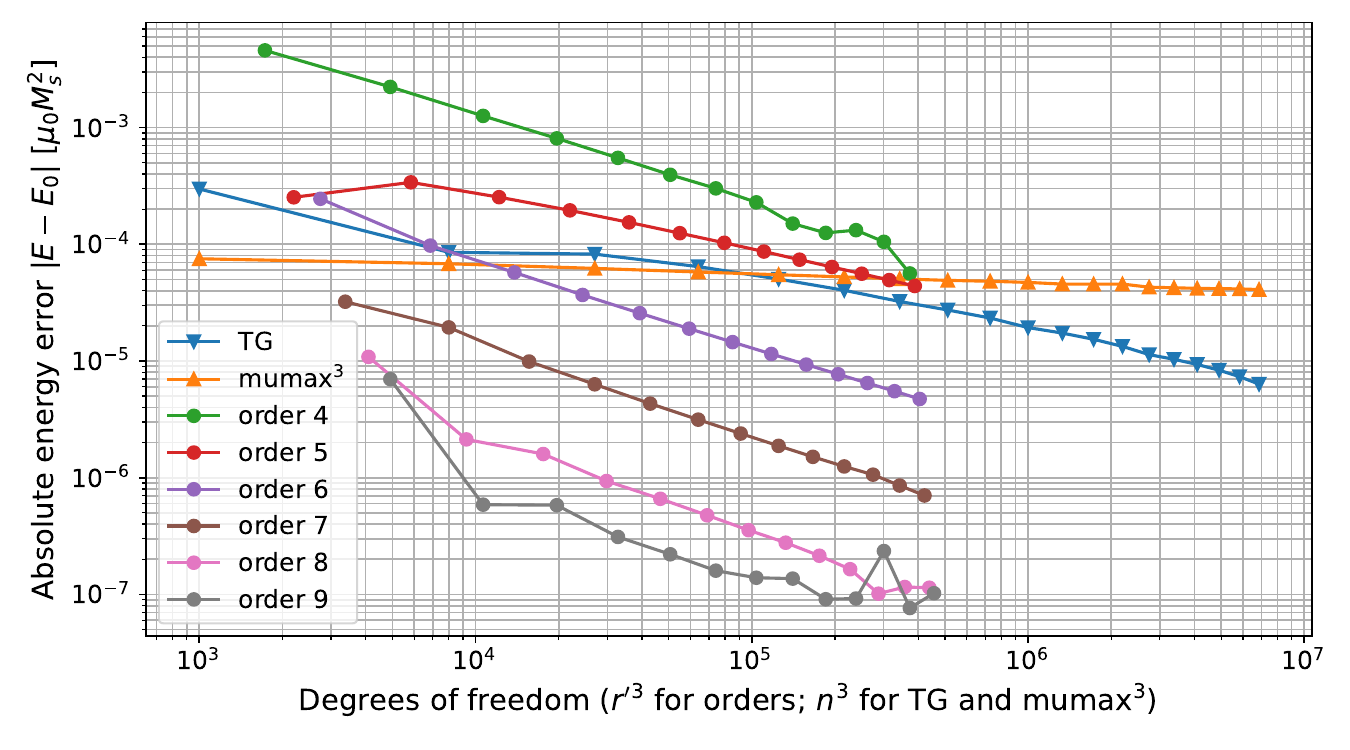}
\caption{\changefirst{Errors of the energy for a homogeneously magnetized cube calculated with TG method, \texttt{mumax$^3$} and Alg.~\ref{alg:ho-stray-field} with $S=100$ for varying field rank $r'$  and order $k$ (super-potential method), and mesh size $n$ (TG and \texttt{mumax$^3$}).}}
\label{fig:energy_uniform_comp}
\end{figure}

\begin{figure}
\center
\includegraphics[width=0.95\textwidth]{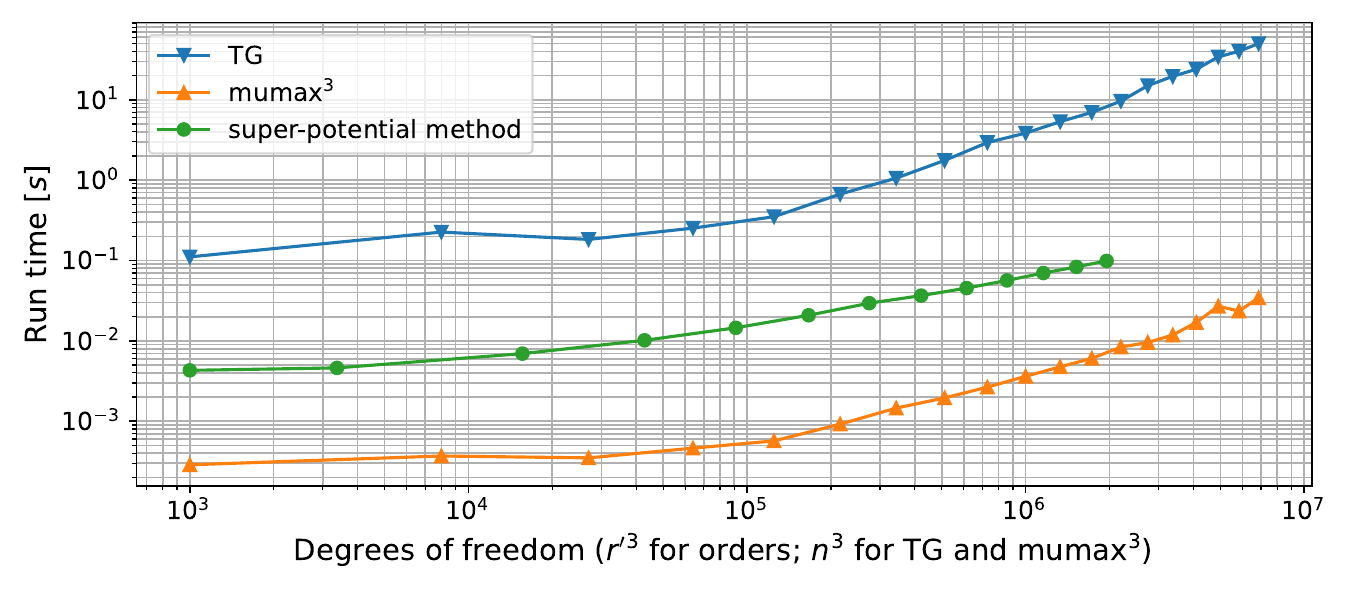}
\caption{\changefirst{Run time for the computation of the energy for a homogeneously magnetized cube calculated with TG method, \texttt{mumax$^3$} and Alg.~\ref{alg:ho-stray-field} with $S=100$ for varying field rank $r'$  and order $k=7$ (super-potential method), and mesh size $n$ (TG and \texttt{mumax$^3$}). Note that TG was computed on CPU, whereas super-potential method and \texttt{mumax$^3$} utilize GPU.}}
\label{fig:run_time}
\end{figure}

We now compute the energy with Alg.~\ref{alg:ho-stray-field} and start with a homogeneously magnetized cube with exact energy $1/6$ [$\mu_0M_s^2$]. The magnetization components are approximated in the functional Tucker tensor form with rank \change{$\bar r^k \equiv \bar r^k_p = 30$} and the same B-spline basis order as the field, leading in all cases to a maximum error of less than $1$e$-13$.  \change{We use $200\times 200\times 200$ quadrature points for all fittings.} 
Fig.~\ref{fig:energy_uniform} shows the absolute error versus rank $r \equiv r_p$ for order $7$. We can observe exponential convergence in the rank.
\changesecond{We expect higher order B-splines to result in a compression in terms of a reduced number of basis functions needed to reach a particular level of accuracy. This is confirmed in Fig.~\ref{fig:energy_uniform_comp}, where we also compared the accuracy to the TG method and \texttt{mumax$^3$}. } 
Tab.~\ref{tab:energy_uniform} shows absolute errors and timings \change{of our super-potential method} for field and energy computation for different order and rank for the field. We observe linear computational scaling in rank $r^\prime$, which is in line with the theoretical cost to calculate the core tensor \eqref{eq:superpotcore}. \change{The computational scaling is also depicted in Fig.~\ref{fig:run_time} for order $k=7$, which also shows the scaling of TG and \texttt{mumax$^3$}}. \change{It should be noted that \texttt{mumax$^3$} uses the demagnetization tensor method which should be exact for this example. We suspect the internal usage of single precision to cause this discrepancy.}

\begin{figure}
\center
\includegraphics[width=0.95\textwidth]{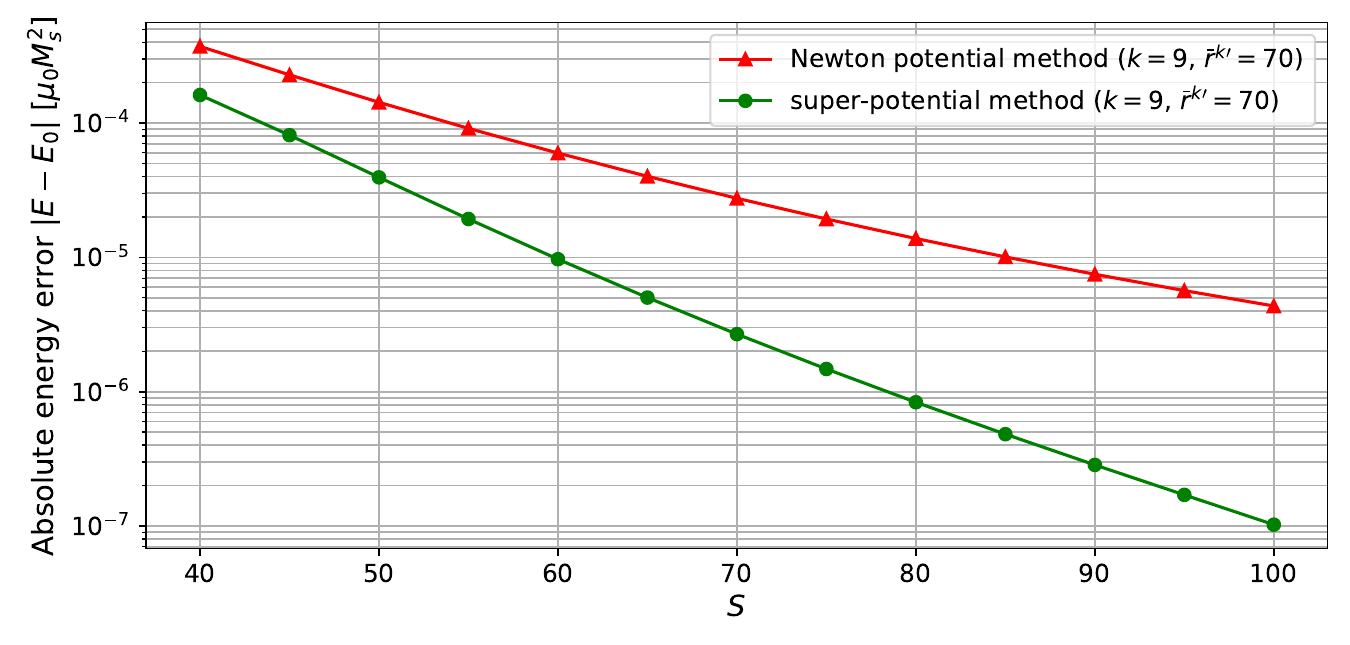}
\caption{\change{Comparison of the Newton potential method versus the super-potential method for a homogeneously magnetized cube.}}
\label{fig:uniform_np_sp_comparison}
\end{figure}

\change{Instead of approximating the super-potential with the GS approximation \eqref{eq:gs_x}, the Newton potential $\mathcal{N}_{\vec m}$ \eqref{eq:newtonpot} can be used with a GS approximation as in \eqref{eq:sin_1_x}. Both can make use of the same sinc quadrature. The stray field can be efficiently computed from the Newton potential by exploiting the same tensor grid structure and using $\vec h = -\nabla(\nabla\cdot\mathcal{N}_{\vec m})$. However, using the Newton potential yields only a cut-off error of $\mathcal{O}(\varepsilon^2)$ compared to $\mathcal{O}(\varepsilon^4)$ for the super-potential. Therefore we expect faster error decay in terms of the expansion rank $S$ for the super-potential method. Fig.~\ref{fig:uniform_np_sp_comparison} also confirms this and shows that the super-potential method is superior. We note that the computational cost for both methods is about the same and both methods show exponential error decay in terms of $S$, where a slight loss of convergence can be observed for large $S$ due to quadrature errors.}

\begin{table}[h!]
\caption{\change{Max. errors in the energy and timings for a homogeneously magnetized cube. B-spline order \(k\), field rank \(\bar r_k^\prime \equiv \bar r_p^{k \prime}\), energy error \(Err\), and timings \(t\) in sec.}}
\label{tab:energy_uniform}
\centering
\setlength{\tabcolsep}{16pt}
\change{
\begin{tabular}{c c c c}
\toprule
\(k\) & \(\bar r_k^\prime\) & \(\mathrm{Err}\) & \(t\) $[s]$ \\
\midrule
4 & 10 & \num{4.586e-03} & \num{4.4e-03} \\
4 & 20 & \num{1.259e-03} & \num{5.3e-03} \\
4 & 40 & \num{3.012e-04} & \num{1.2e-02} \\
\midrule
5 & 10 & \num{2.526e-04} & \num{4.9e-03} \\
5 & 20 & \num{2.533e-04} & \num{6.0e-03} \\
5 & 40 & \num{1.029e-04} & \num{1.3e-02} \\
\midrule
6 & 10 & \num{2.453e-04} & \num{5.1e-03} \\
6 & 20 & \num{5.733e-05} & \num{7.0e-03} \\
6 & 40 & \num{1.450e-05} & \num{1.4e-02} \\
\midrule
7 & 10 & \num{3.219e-05} & \num{5.2e-03} \\
7 & 20 & \num{9.890e-06} & \num{7.8e-03} \\
7 & 40 & \num{2.389e-06} & \num{1.4e-02} \\
\midrule
8 & 10 & \num{1.085e-05} & \num{5.2e-03} \\
8 & 20 & \num{1.594e-06} & \num{7.9e-03} \\
8 & 40 & \num{3.549e-07} & \num{1.5e-02} \\
\midrule
9 & 10 & \num{6.986e-06} & \num{6.1e-03} \\
9 & 20 & \num{5.805e-07} & \num{7.7e-03} \\
9 & 40 & \num{1.388e-07} & \num{1.5e-02} \\
\bottomrule
\end{tabular}
}
\end{table}

\subsubsection{\change{Flower and vortex state}}

\begin{figure}[ht]
  \centering

  % First subfigure
  \begin{subfigure}{0.5\textwidth}
    \centering
    \includegraphics[width=1.2\linewidth]{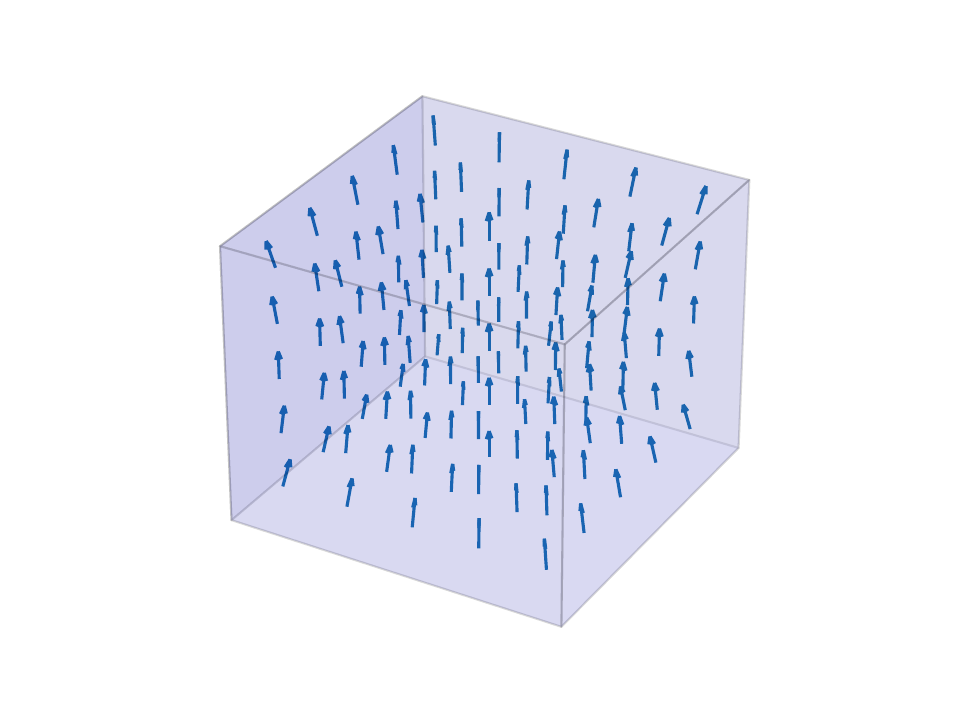}
    \caption{Flower state}
    \label{fig:flower}
  \end{subfigure}%
  % Second subfigure
  \begin{subfigure}{0.5\textwidth}
    \centering
    \includegraphics[width=1.2\linewidth]{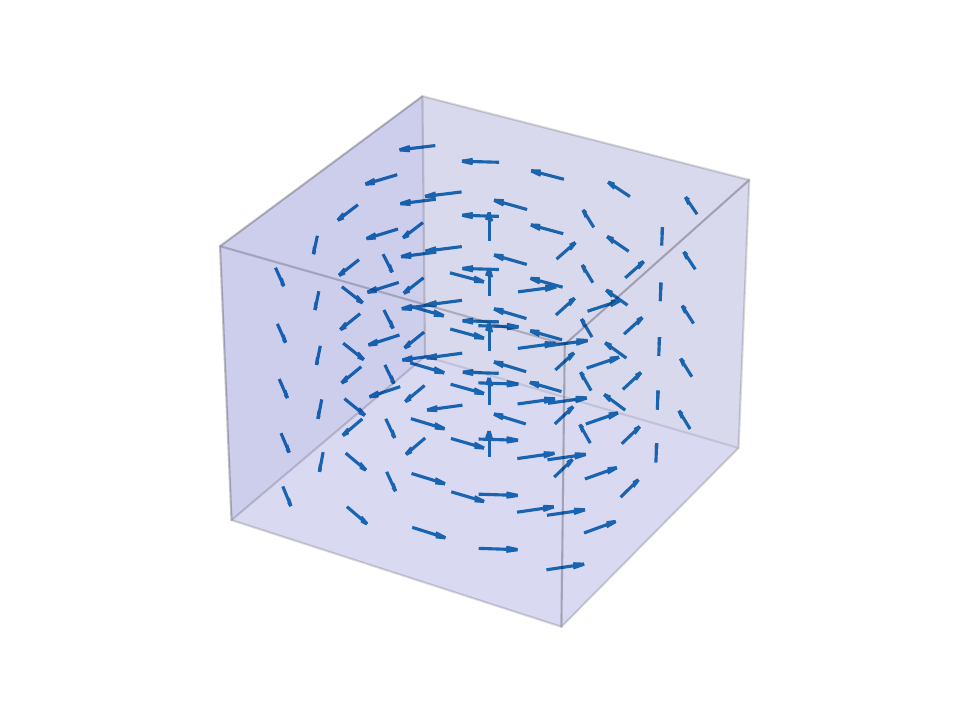}
    \caption{Vortex state}
    \label{fig:vortex}
  \end{subfigure}
  % Overall figure caption
  \caption{Visualization of two different magnetic configurations in the unit cube.}
  \label{fig:flower_vortex}
\end{figure}

\changefirst{Next, we compute the energy of the flower state and the vortex state in the unit cube with reference value $1.528$e-$01$ [$\mu_0M_s^2$] and $2.189$e-$02$ [$\mu_0M_s^2$], respectively, from the DM method  on a $40 \times 40 \times 40$ grid \cite{abert2013numerical}. Fig.~\ref{fig:flower_vortex} visualizes these magnetic configurations. The flower state is fitted as a functional Tucker tensor of rank $\bar r^k \equiv \bar r^k_p = 40$ and the vortex state as a functional Tucker tensor of rank $\bar r^k \equiv \bar r^k_p = 80$. For all fitting procedures $300\times300\times300$ quadrature nodes are used on a tensor grid. Tab.~\ref{tab:energy_flower_vortex} shows the results for different order (for field and magnetization) and rank of the field together with numerical results for the TG method and \texttt{mumax$^3$}.}

\begin{table}[h!]
\caption{\changefirst{Energy for a flower state and a vortex state in the unit cube computed with super-potential method, TG method and \texttt{mumax$^3$}. B-spline order \(k\), field rank \(\bar r^k{'} \equiv \bar r_p^k{'}\), mesh size $n$ for TG and \texttt{mumax$^3$}, energy \(e\) $[\mu_0 M_s^2]$ and timings $[s]$. The reference values from a DM method are $1.528$e-$01$ [$\mu_0M_s^2$] and $2.189$e-$02$ [$\mu_0M_s^2$], respectively. Note that TG was computed on CPU, whereas super-potential method and \texttt{mumax$^3$} utilize GPU.}} 
\label{tab:energy_flower_vortex}
\centering
\setlength{\tabcolsep}{13pt}
\changefirst{
\begin{tabular}{c c c c c c}
\toprule
&& \multicolumn{2}{c}{flower state} &\multicolumn{2}{c}{vortex state}\\
\(k\) & \(\bar r^k{'}\) & \(e\) $[\mu_0 M_s^2]$  & $t \, [s]$& \(e\) $[\mu_0 M_s^2]$ & $t\, [s]$\\
\midrule
4 & 10 & \num{1.48120e-01} & \num{5.4e-03} & \num{1.82836e-02} & \num{1.4e-02} \\
4 & 20 & \num{1.51520e-01} & \num{8.3e-03} & \num{2.08391e-02} & \num{2.2e-02} \\
4 & 40 & \num{1.52504e-01} & \num{1.6e-02} & \num{2.15361e-02} & \num{4.0e-02} \\
4 & 80 & \num{1.52753e-01} & \num{5.1e-02} & \num{2.17331e-02} & \num{1.1e-01} \\
\midrule
5 & 10 & \num{1.53067e-01} & \num{6.4e-03} & \num{2.11988e-02} & \num{1.5e-02} \\
5 & 20 & \num{1.53060e-01} & \num{9.1e-03} & \num{2.18939e-02} & \num{2.2e-02} \\
5 & 40 & \num{1.52906e-01} & \num{1.8e-02} & \num{2.18538e-02} & \num{3.9e-02} \\
5 & 80 & \num{1.52837e-01} & \num{5.7e-02} & \num{2.18184e-02} & \num{1.2e-01} \\
\midrule
6 & 10 & \num{1.53065e-01} & \num{6.1e-03} & \num{2.38648e-02} & \num{1.5e-02} \\
6 & 20 & \num{1.52863e-01} & \num{9.9e-03} & \num{2.18415e-02} & \num{2.3e-02} \\
6 & 40 & \num{1.52816e-01} & \num{1.9e-02} & \num{2.18078e-02} & \num{3.9e-02} \\
6 & 80 & \num{1.52805e-01} & \num{5.7e-02} & \num{2.17992e-02} & \num{1.1e-01} \\
\midrule
7 & 10 & \num{1.52849e-01} & \num{7.0e-03} & \num{2.23807e-02} & \num{1.5e-02} \\
7 & 20 & \num{1.52813e-01} & \num{1.1e-02} & \num{2.18032e-02} & \num{2.6e-02} \\
7 & 40 & \num{1.52804e-01} & \num{2.0e-02} & \num{2.17980e-02} & \num{4.6e-02} \\
7 & 80 & \num{1.52801e-01} & \num{5.9e-02} & \num{2.17968e-02} & \num{1.2e-01} \\
\midrule
8 & 10 & \num{1.52819e-01} & \num{7.5e-03} & \num{1.69230e-02} & \num{1.6e-02} \\
8 & 20 & \num{1.52804e-01} & \num{1.1e-02} & \num{2.17974e-02} & \num{2.4e-02} \\
8 & 40 & \num{1.52802e-01} & \num{2.1e-02} & \num{2.17964e-02} & \num{4.4e-02} \\
8 & 80 & \num{1.52801e-01} & \num{6.2e-02} & \num{2.17964e-02} & \num{1.1e-01} \\
\midrule\midrule
&$n$ & & & & \\
\midrule
\multirow{5}{*}{\rotatebox{90}{TG}} & 40 & \num{1.52896e-01} & \num{2.9e-01} & \num{2.19328e-02} & \num{2.9e-01} \\
&80 & \num{1.52836e-01} & \num[print-zero-exponent=true]{2.1e+00} & \num{2.18420e-02} & \num[print-zero-exponent=true]{2.1e+00} \\
&120 & \num{1.52820e-01} & \num[print-zero-exponent=true]{9.6e+00} & \num{2.18195e-02} & \num{1.0e+01} \\
&160 & \num{1.52812e-01} & \num{2.7e+01} & \num{2.18104e-02} & \num{3.3e+01} \\
&200 & \num{1.52807e-01} & \num{6.5e+01} & \num{2.18050e-02} & \num{6.0e+01} \\
\midrule\midrule
\multirow{4}{*}{\rotatebox{90}{\texttt{mumax$^3$}}}&40 & \num{1.52764e-01} & \num{5.0e-04} & \num{2.18448e-02} & \num{4.3e-04} \\
&80 & \num{1.52757e-01} & \num{2.0e-03} & \num{2.17768e-02} & \num{2.0e-03} \\
&120 & \num{1.52759e-01} & \num{6.1e-03} & \num{2.17654e-02} & \num{6.3e-03} \\
&160 & \num{1.52757e-01} & \num{1.7e-02} & \num{2.17622e-02} & \num{1.7e-02} \\
&200 & \num{1.52759e-01} & \num{3.3e-02} & \num{2.17604e-02} & \num{3.3e-02} \\
\bottomrule
\end{tabular}
}
\end{table}

\subsubsection{\change{Vortex state in a thin film}}
\change{Next}, we compute \changefirst{the energy of} the vortex state in the $1\times 1 \times 0.1$ thin film from \cite{abert2013numerical} with reference value $1.569$e$-03$ [$\mu_0 M_s^2$] from the DM method on a grid of size $80 \times 80 \times 8$. We used multilinear rank $(40,40,10)$ for the magnetization. \change{We use $300\times 300\times 75$ quadrature points for all fittings.} Tab.~\ref{tab:energy_vortex_thinfilm} shows the results for different order (for field and magnetization) and rank of the field.

\begin{table}[h!]
\caption{\changefirst{Computed} energy for a thin film \change{vortex} state of domain $1\times 1\times 0.1$. B-spline order \(k\), \change{multilinear field rank \((\bar r_1^{k\prime}, \bar r_2^{k\prime},\bar r_3^{k\prime})\)}, \change{mesh size $(n_1,n_2,n_3)$}, energy \(e\) $[\mu_0 M_s^2]$ \change{and timings $t$ $[s]$}.} 
\label{tab:energy_vortex_thinfilm}
\centering
\setlength{\tabcolsep}{16pt}
\change{
\begin{tabular}{c c c c}
\toprule
\(k\) & \((\bar r_1^k{'}, \bar r_2^k{'}, \bar r_3^k{'})\) & \(e\) [$\mu_0 M_s^2$] & $t\,[s]$\\
\midrule
5 & $20\times20\times5$ & \num{1.53173e-03} & \num{4.4e-03} \\
5 & $40\times40\times10$ & \num{1.56705e-03} & \num{7.7e-03} \\
5 & $60\times60\times15$ & \num{1.56995e-03} & \num{1.2e-02} \\
\midrule
6 & $20\times20\times5$ & \num{1.56445e-03} & \num{4.8e-03} \\
6 & $40\times40\times10$ & \num{1.56947e-03} & \num{7.0e-03} \\
6 & $60\times60\times15$ & \num{1.56854e-03} & \num{1.1e-02} \\
\midrule
7 & $20\times20\times5$ & \num{1.56906e-03} & \num{6.5e-03} \\
7 & $40\times40\times10$ & \num{1.56767e-03} & \num{9.6e-03} \\
7 & $60\times60\times15$ & \num{1.56749e-03} & \num{1.7e-02} \\
\midrule
8 & $20\times20\times5$ & \num{1.56779e-03} & \num{6.9e-03} \\
8 & $40\times40\times10$ & \num{1.56711e-03} & \num{1.0e-02} \\
8 & $60\times60\times15$ & \num{1.56730e-03} & \num{1.7e-02} \\
\midrule\midrule
& $(n_1,n_2,n_3)$ &  & \\
\midrule
\multirow{4}{*}{\rotatebox{90}{TG}} & $40\times 40 \times 10 $ & \num{1.55549e-03} & \num{1.9e-01} \\
&$80\times 80 \times 20 $ & \num{1.56660e-03} & \num{5.3e-01} \\
&$120\times 120 \times 30 $ & \num{1.56730e-03} & \num[print-zero-exponent=true]{1.7e0}\\
&$160\times 160 \times 40 $ & \num{1.56664e-03} & \num[print-zero-exponent=true]{5.0e0} \\
&$200\times 200 \times 50 $ & \num{1.56565e-03} & \num{1.1e+01} \\
\midrule\midrule
\multirow{4}{*}{\rotatebox{90}{\texttt{mumax$^3$}}} & $40\times 40 \times 10 $ & \num{1.57970e-03} & \num{3.6e-04} \\
&$80\times 80 \times 20 $ & \num{1.56828e-03} & \num{6.4e-04} \\
&$120\times 120 \times 30 $ & \num{1.56497e-03} & \num{1.8e-03} \\
&$160\times 160 \times 40 $ & \num{1.56343e-03} & \num{3.6e-03} \\
&$200\times 200 \times 50 $ & \num{1.56254e-03} & \num{8.8e-03} \\
\bottomrule
\end{tabular}
}
\end{table}

\subsubsection{\change{Two layer example}}
\begin{figure}[ht]
    \centering
    \includegraphics[width=0.6\linewidth]{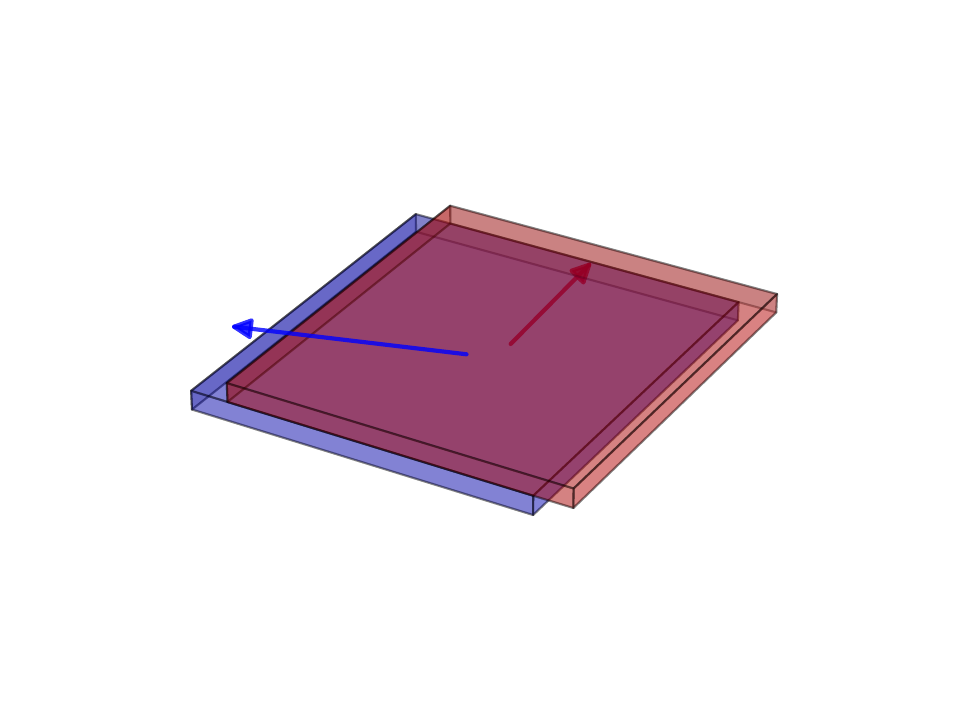}
    \caption{\change{Two magnetized thin layers with homogeneous magnetization.}}
    \label{fig:two_layer}
\end{figure}
\change{Our implementation allows the evaluation of the super-potential on multiple tensor product domains. Assume we have two tensor product domains $\Omega_1$ and $\Omega_2$ with magnetization $\vec m_1$ and $\vec m_2$, respectively, then we can compute the super-potential by adding the contributions of each domain}
\change{
\begin{equation}\label{eq:superpot_multidomain}
    \vec u^{\Omega_1}_{\vec m} = \vec u_{\vec m_1}^{\Omega_1} + \vec u_{\vec m_2}^{\Omega_1} \qquad
    \vec u^{\Omega_2}_{\vec m} = \vec u_{\vec m_1}^{\Omega_2} + \vec u_{\vec m_2}^{\Omega_2},
\end{equation}
where $\vec u_{\vec m_j}^{\Omega_i}$ denotes the super-potential evaluated in $\Omega_i$ and induced by the magnetization configuration $\vec m_j$ in $\Omega_j$. Each super-potential in \eqref{eq:superpot_multidomain} can be computed individually with Alg.~\ref{alg:ho-stray-field} avoiding large errors near the boundaries. 
}
\change{For $m$ domains, $m^2$ super-potential fitting procedures are required. However, it should be noted that $S$ could be reduced for domains which are further apart, which would save on computational cost.}

\change{For our final example we demonstrate the super-potential method for a thin film with two layers. The top layer $\Omega_1=[-0.4, 0.5]\times[-0.5, 0.5]\times[0, 0.05]$ has a uniform magnetization in the $[0.4, 1, 0.6]^T$ direction and the bottom layer $\Omega_2=[-0.5, 0.4]\times[-0.5, 0.5]\times[-0.05, 0]$ has a uniform magnetization in the $[-1, -0.3, 0.0]^T$ direction. Additionally, we use $M_{s,\Omega_1} = 1$ in $\Omega_1$ and $M_{s, \Omega_2}=2$ in $\Omega_2$. Fig~\ref{fig:two_layer} depicts this setting.}

\change{For each domain we use $k=8$ and a multilinear rank $(\bar r_1^k,\bar r_2^k,\bar r_3^k) = (30, 30, 4)$ to fit the magnetization as well as the rank $(\bar r_1^k{'},\bar r_2^k{'},\bar r_3^k{'}) = (40, 40, 5)$ for the field. 
Further, we use $(200\times200\times25)$ quadrature nodes per domain. We find, for $S=100$ an energy $e=\num{1.08631e-2}$. We used \texttt{mumax$^3$} with mesh $(100\times100\times10)$  to verify this result and find an energy of $\num{1.05553e-2}$. This shows that our method can also be applied to more complex tensor product domains.}

\section{Conclusion and Discussion}
The proposed method extends the tensor grid approach from \cite{exl2012fast} to higher-order B-spline basis expansions. While the original method employed a tensor product of indicator functions (corresponding to B-splines of order $k=1$), our approach generalizes this by using B-splines of \change{order $k>3$ - an assumption guaranteeing smooth derivatives}. This results in a natural compression of the core tensor, compared to the full tensor grid core, due to the smoother and more compact representation afforded by higher-order B-splines.
Numerical tests confirm that our method achieves higher accuracy while maintaining relatively low multilinear rank. The computational scaling is effectively (quasi-)linear: linear with respect to the number of degrees of freedom in the magnetization (i.e., the knot grid), and linear in the one-dimensional rank components of the field. A further computational advantage is that the ranks of both the magnetization and the field can be adaptively controlled through an $L^2$-loss fitting procedure.

The use of a \textit{super-potential} formulation for the field introduces a smooth convolution kernel, which can be accurately and separably approximated. This allows us to exploit the tensor product structure of the domain for significant computational gains. Central to our method is fitting the super-potential as a \textit{functional Tucker tensor}, using a multilinear tensor product extension of the extreme learning machine methodology. Notably, this avoids the singular integrals that would arise from direct or automatic differentiation.

The higher-order scheme offers a range of promising applications, particularly due to its mesh-free and continuous formulation of the field. Complete micromagnetic simulations, including multiple energy contributions, could be carried out entirely within the functional Tucker tensor framework. This approach also allows for more natural and accurate treatment of boundary conditions, which are often challenging in traditional discrete settings. Moreover, both Landau–Lifshitz–Gilbert (LLG) dynamics and eigenmode analysis appear feasible within a quasi-analytical, mesh-free computational framework, potentially improving both accuracy and efficiency.
\changesecond{Our approach can be adapted to other types of nonlocal potentials, such as the Coulomb potential \cite{exl2025nearfieldfree}}, which appears in a wide range of mathematical models across quantum physics, quantum chemistry, materials science, plasma physics, and computational biology. In these contexts, the method is also compatible with Fourier bases, which offer certain advantages—most notably, their linearity stemming from the orthogonality of basis functions. In such cases, no $L^2$-loss fitting procedure is required.

In summary, the proposed framework demonstrates how higher-order functional tensor approximations can significantly enhance both the accuracy and efficiency of micromagnetic field computations. By combining smooth basis functions, adaptive low-rank representations, and a mesh-free formulation, the method offers a scalable and versatile alternative to traditional grid-based approaches. Given its generality and compatibility with other basis systems such as Fourier functions, the methodology opens up promising avenues for further research and application in a variety of fields involving nonlocal interactions. Future work may focus on extending this approach to dynamic simulations, coupling with other physical phenomena, and integration into large-scale computational platforms.

\newpage
\section*{Acknowledgements}
\noindent Financial support by the Austrian Science Fund (FWF) via project ”Data-driven Reduced Order Approaches for Micromagnetism (Data-ROAM)” (Grant-DOI: 10.55776/PAT7615923) and project ”Design of Nanocomposite Magnets by
Machine Learning (DeNaMML)” (Grant-DOI: 10.55776/P35413) is gratefully acknowledged. The authors acknowledge the University of Vienna research platform MMM Mathematics - Magnetism - Materials. The computations were partly achieved by using the Vienna Scientific Cluster (VSC) via the funded projects No. 71140, 71952 and 72862.
This research was funded in whole or in part by the Austrian Science Fund (FWF) [10.55776/PAT7615923, 10.55776/P35413]. For the purpose of Open Access, the authors have applied a CC BY public copyright license to any Author Accepted Manuscript (AAM) version arising from this submission. 
%%%%%%%%%%%%%%%%%%%%

%\bibliographystyle{abbrv}
%\bibliography{bibref}

\end{document}